\documentclass[3p,times,procedia]{elsarticle}
\usepackage{nupha_ecrc}
\usepackage[utf8]{inputenc}

\volume{00}

\firstpage{1}

\journalname{Nuclear Physics A}

\runauth{}

\jid{nupha}

\jnltitlelogo{Nuclear Physics A}

\usepackage{amssymb}

\usepackage[figuresright]{rotating}

\begin{document}

\begin{frontmatter}



\dochead{XXVIIIth International Conference on Ultrarelativistic Nucleus-Nucleus Collisions\\ (Quark Matter 2019)}

\title{Bulk properties and multi-particle correlations\\ in large and small systems}


\author[1]{Bj\"orn Schenke}
\author[2,3]{Chun Shen}
\author[1]{Prithwish Tribedy}

\address[1]{Physics Department, Brookhaven National Laboratory, Upton, NY 11973, USA}
\address[2]{Department of Physics and Astronomy, Wayne State University, Detroit, Michigan 48201, USA}
\address[3]{RIKEN BNL Research Center, Brookhaven National Laboratory, Upton, NY 11973, USA}

\begin{abstract}
Charged particle production is calculated in a hybrid framework consisting of the IP-Glasma initial state, \textsc{Music} viscous relativistic fluid dynamics, and the UrQMD microscopic hadronic cascade. 
Using one set of parameters, we compute observables for a large variety of collision systems. We compare to experimental data in p+p, p+Pb, Xe+Xe, Au+Au and Pb+Pb collisions at various energies, and make predictions for potential O+O runs at RHIC and LHC.
\end{abstract}

\begin{keyword}
heavy ion collisions \sep hydrodynamics \sep multiparticle correlations \sep small system collisions


\end{keyword}

\end{frontmatter}


\section{Introduction}
\label{sec:intro}
Bulk particle production and multi-particle correlations in heavy ion collisions are generally well described in event-by-event frameworks that evolve an initial spatially dependent energy momentum tensor according to hydrodynamics and describe the low temperature regime using hadronic cascades \cite{Gale:2013da,Petersen:2014yqa}. In this work we explore how well the framework consisting of the IP-Glasma initial state \cite{Schenke:2012wb,Schenke:2012hg}, \textsc{Music} fluid dynamics \cite{Schenke:2010nt,Schenke:2010rr,Schenke:2011bn}, and UrQMD hadronic cascade \cite{Bass:1998ca,Bleicher:1999xi} can describe experimental data over a wide range of (transverse) system size, collision energy, and charged hadron multiplicity, using one fixed set of parameters.
We make predictions for $v_2\{2\}$ and $v_3\{2\}$ in O+O collisions at RHIC and LHC energies.

\section{Model}
\label{sec:model}
\begin{figure}[tb]
  \centering
  \includegraphics[width=0.6\textwidth]{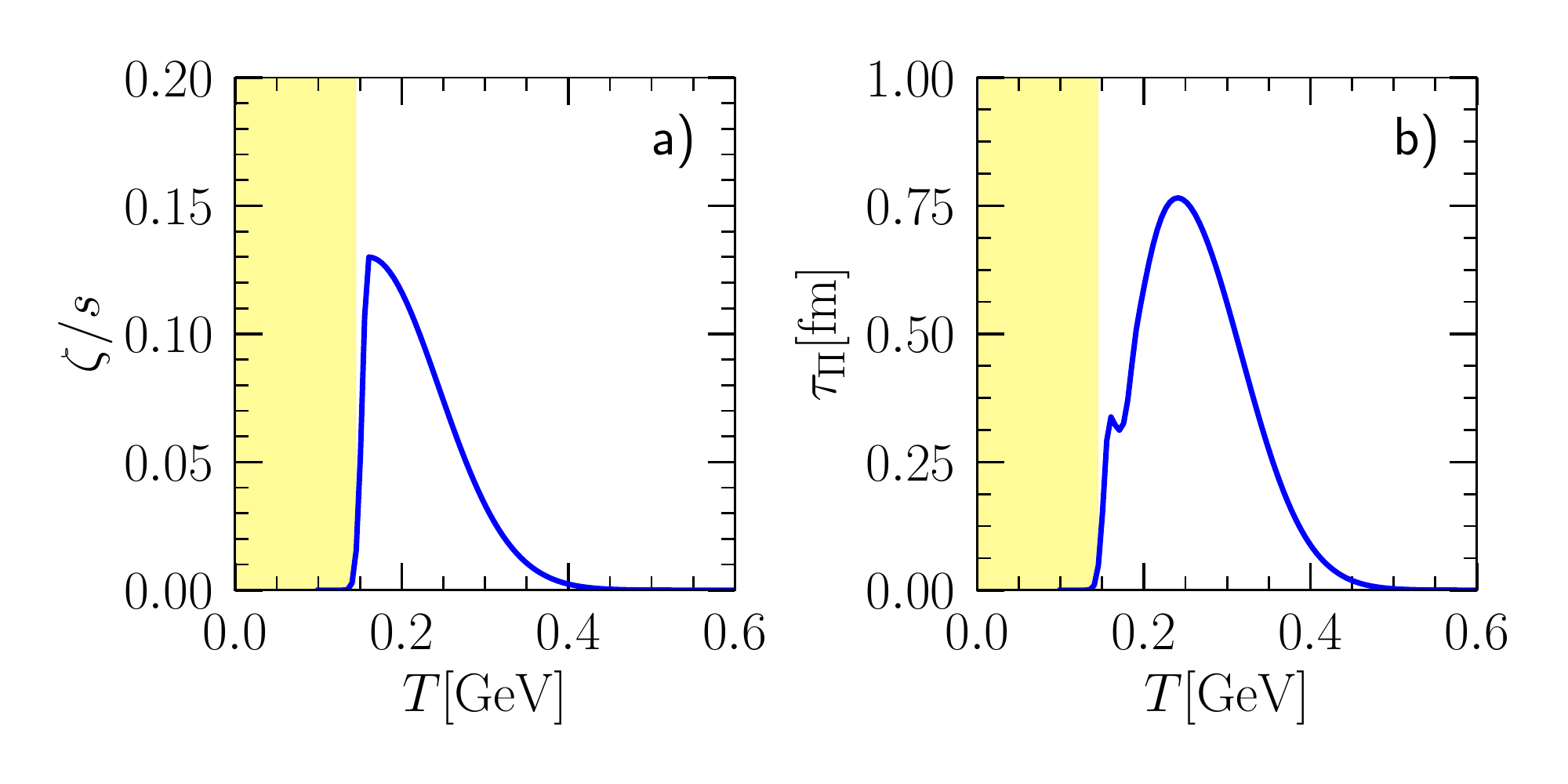}\vspace{-0.5cm}
  \caption{Bulk viscosity over entropy density (a) and bulk relaxation time (b) vs. temperature. \label{fig:bulk}}
\end{figure}

The model framework is described in \cite{Mantysaari:2017cni,Schenke:2019ruo,Schenke:2019pmk}. The IP-Glasma initial state includes fluctuations of nucleon positions, three sub-nucleon hotspots per nucleon, as well as color charge fluctuations. We switch to the hydrodynamic simulation at $\tau_{\rm init}=0.4\,{\rm fm}/c$, and use a constant shear viscosity to entropy density ratio $\eta/s=0.12$ and a temperature dependent bulk viscosity to entropy density ratio $\zeta/s$ shown in Fig.\,\ref{fig:bulk} together with the bulk relaxation time $\tau_\Pi$. Below the switching temperature of $T_{\rm sw}=145\,{\rm MeV}$, we evolve the system within UrQMD. 
More details on the model can be found in \cite{Schenke:2019pmk}.

\section{Results}
We present results for bulk observables for a large variety of collision systems and energies. We do not vary any parameters other than the target and projectile species and collision energy in the IP-Glasma calculation. This allows for robust predictions of multiplicities, mean transverse momentum, and anisotropic flow for a variety of systems, including O+O collisions to be performed in the future.

\begin{figure}[tb]
  \centering
  \includegraphics[width=0.45\textwidth]{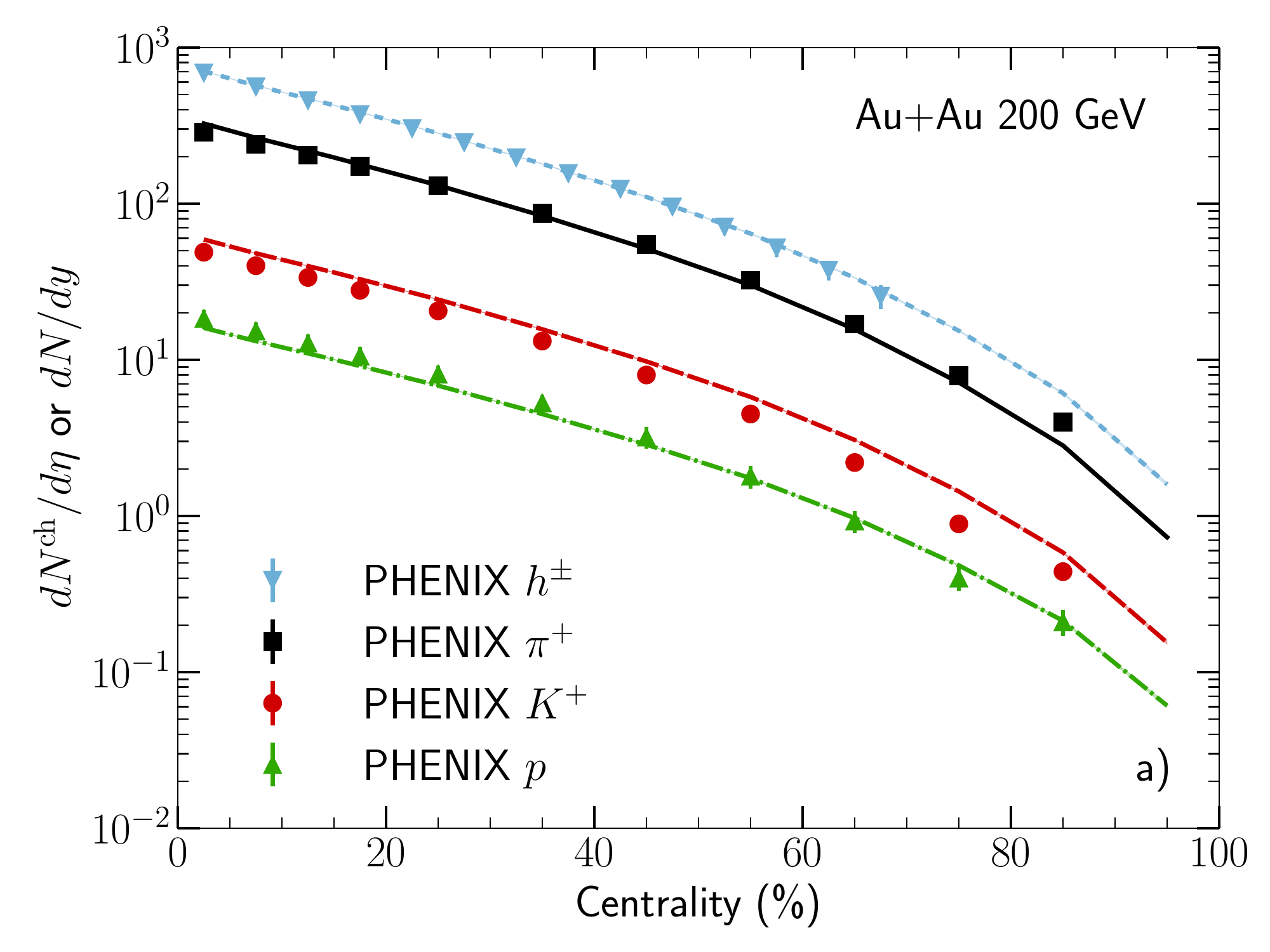}\includegraphics[width=0.45\textwidth]{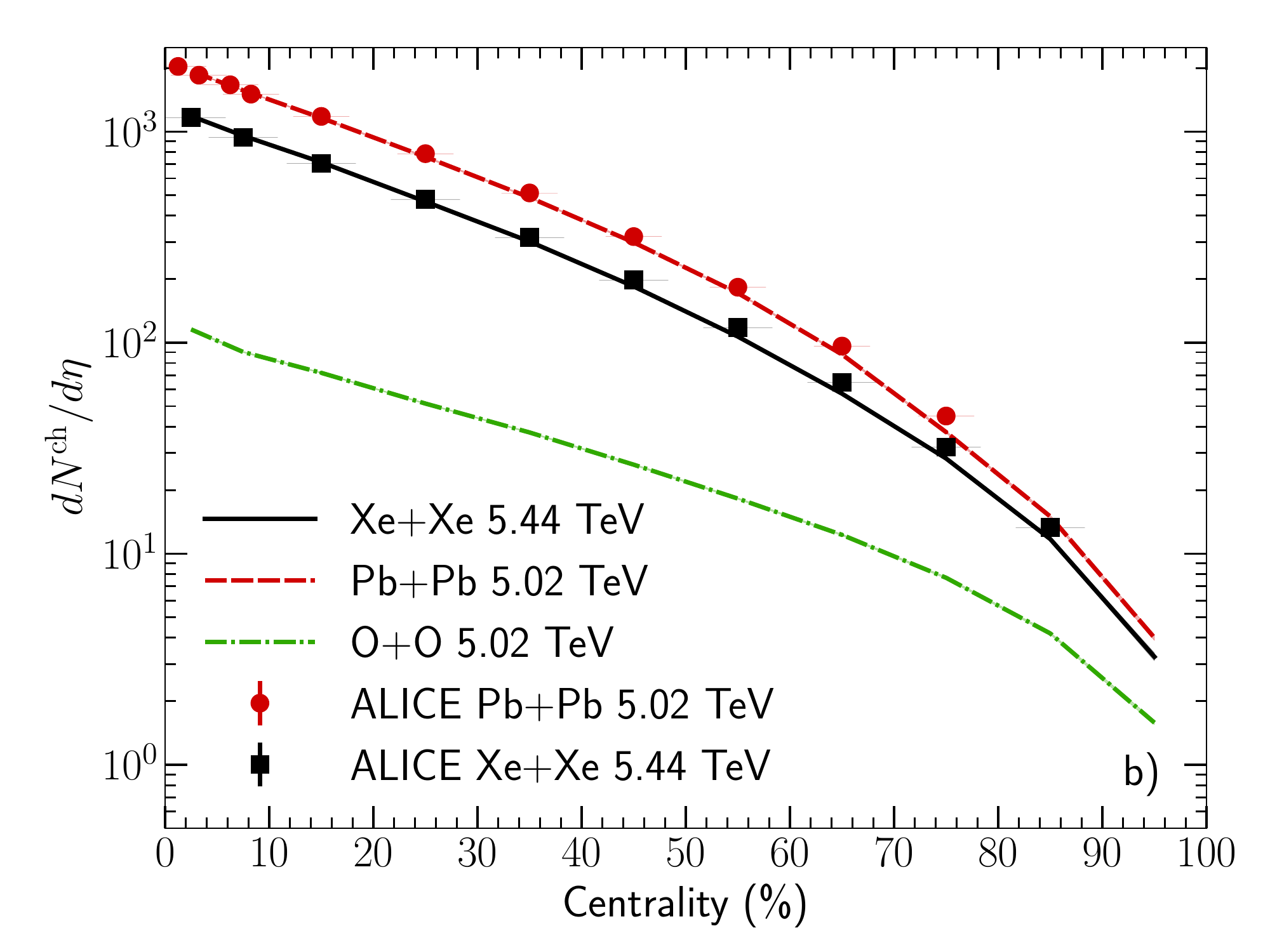}
  \caption{Charged hadron and identified particle multiplicities vs.~centrality in Au+Au collisions at RHIC (a), and charged hadron multiplicities in O+O, Xe+Xe, and Pb+Pb collisions at LHC (b). Experimental data from PHENIX \cite{Adler:2003cb} and ALICE \cite{Adam:2015ptt,Acharya:2018hhy}. \label{fig:mult}}
\end{figure}

\begin{figure}[tb]
  \centering
  \includegraphics[width=0.45\textwidth]{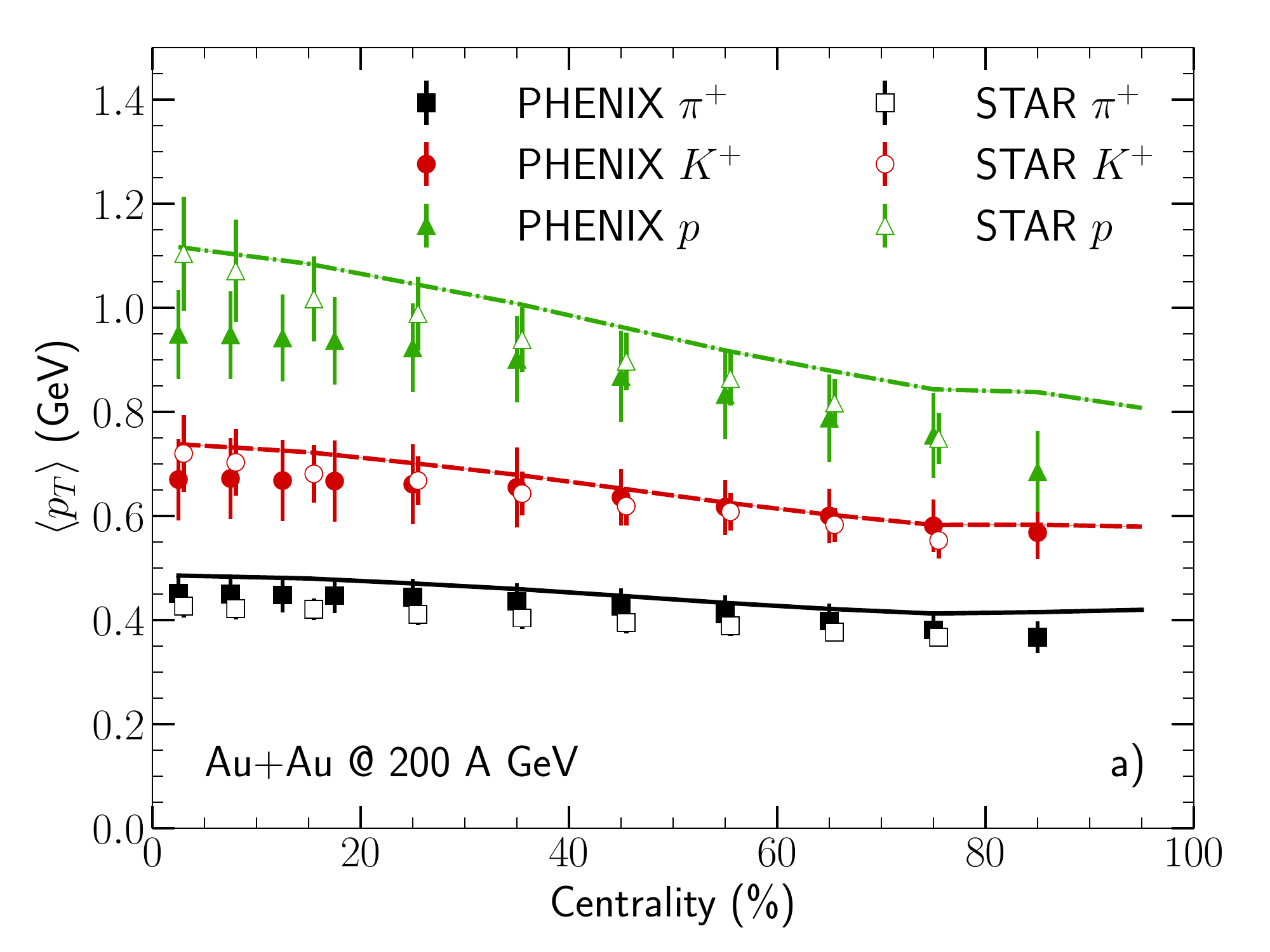}\includegraphics[width=0.45\textwidth]{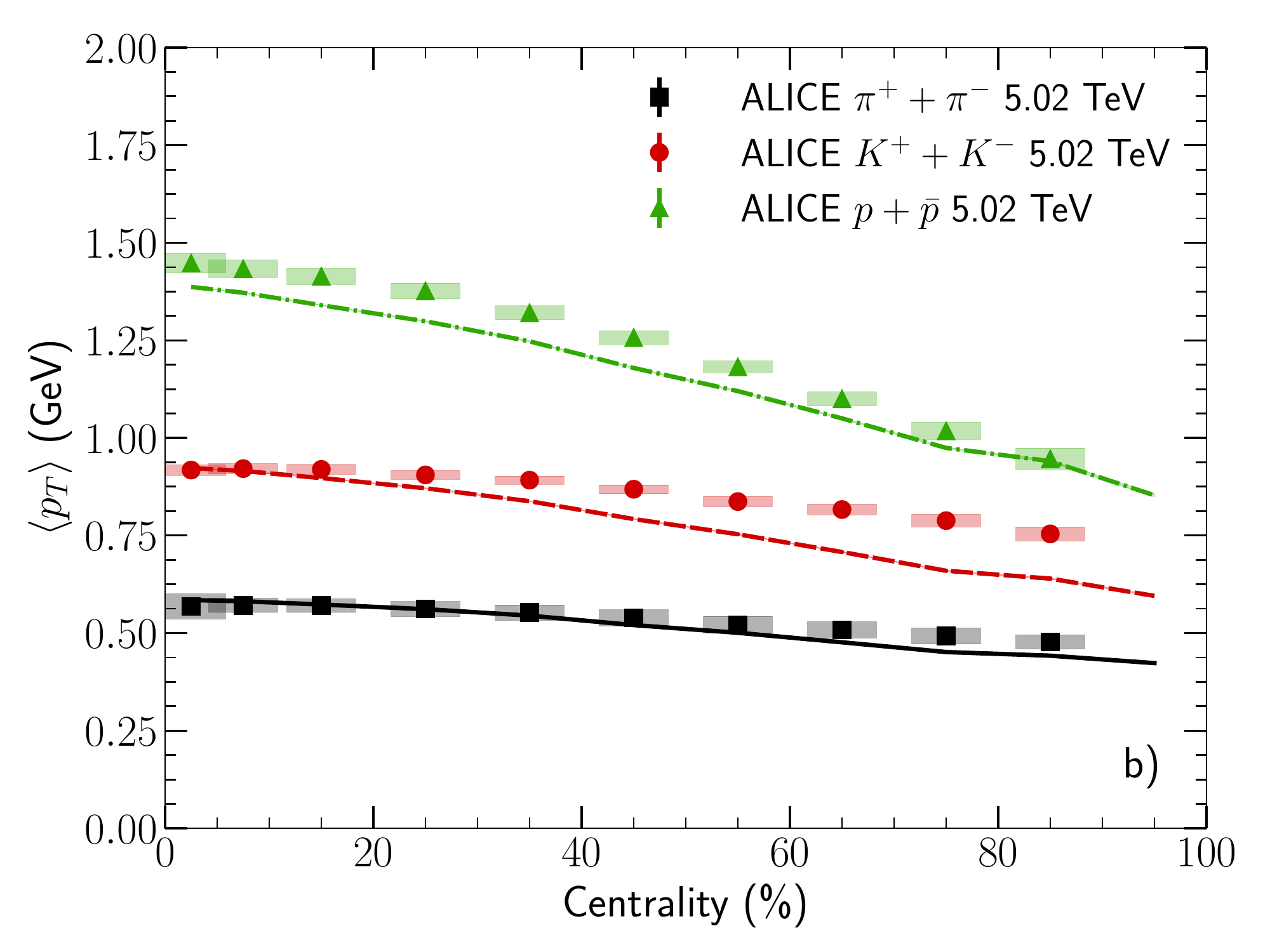}
  \caption{Identified particle mean transverse momentum vs.~centrality in Au+Au collisions at RHIC (a) and Pb+Pb collisions at LHC (b). Experimental data from PHENIX \cite{Adler:2003cb}, STAR \cite{Abelev:2008ab}, and ALICE \cite{Acharya:2019yoi}. \label{fig:pt}}
\end{figure}

\begin{figure}[tb]
  \centering
  \includegraphics[width=0.45\textwidth]{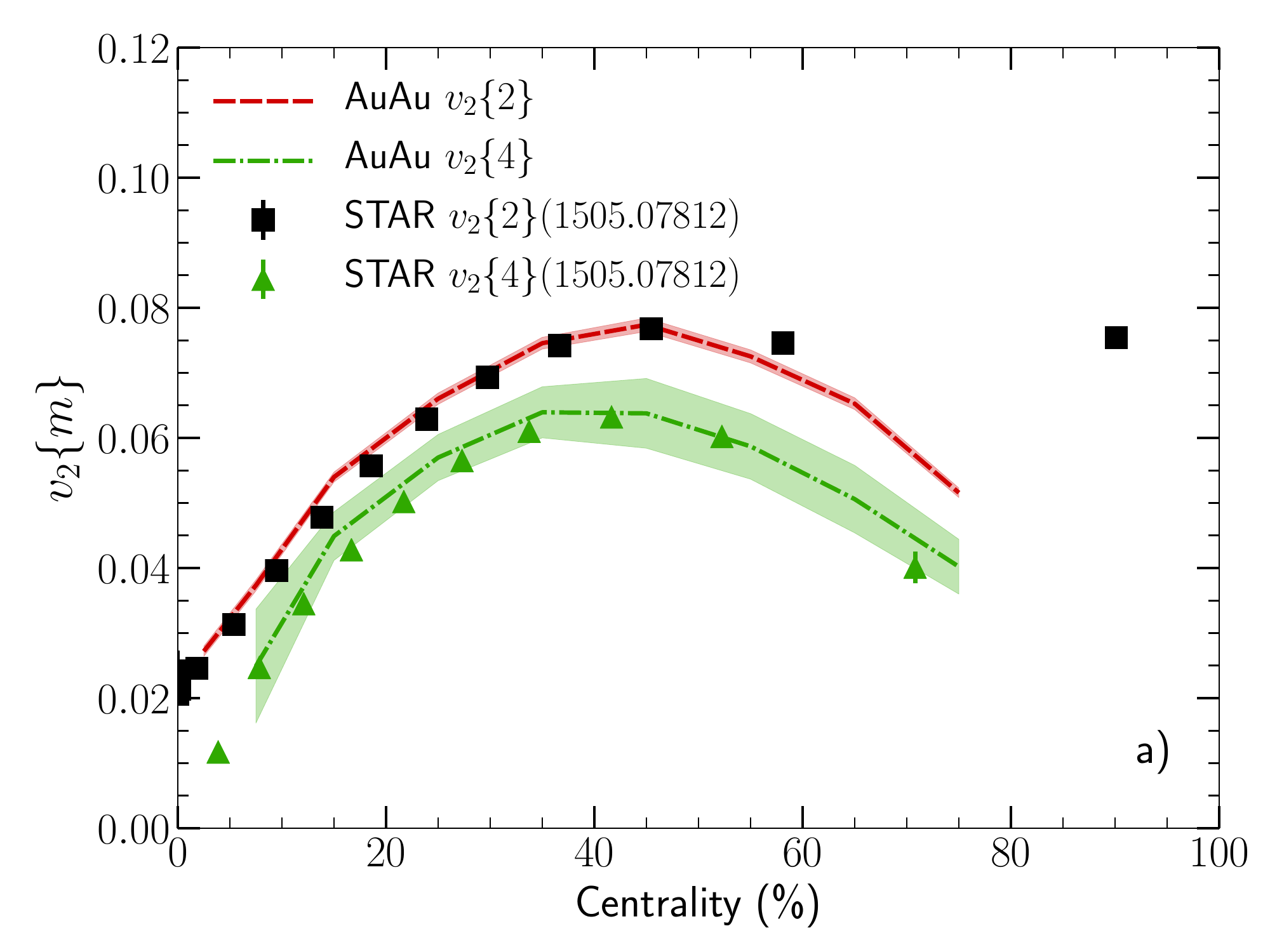}\includegraphics[width=0.45\textwidth]{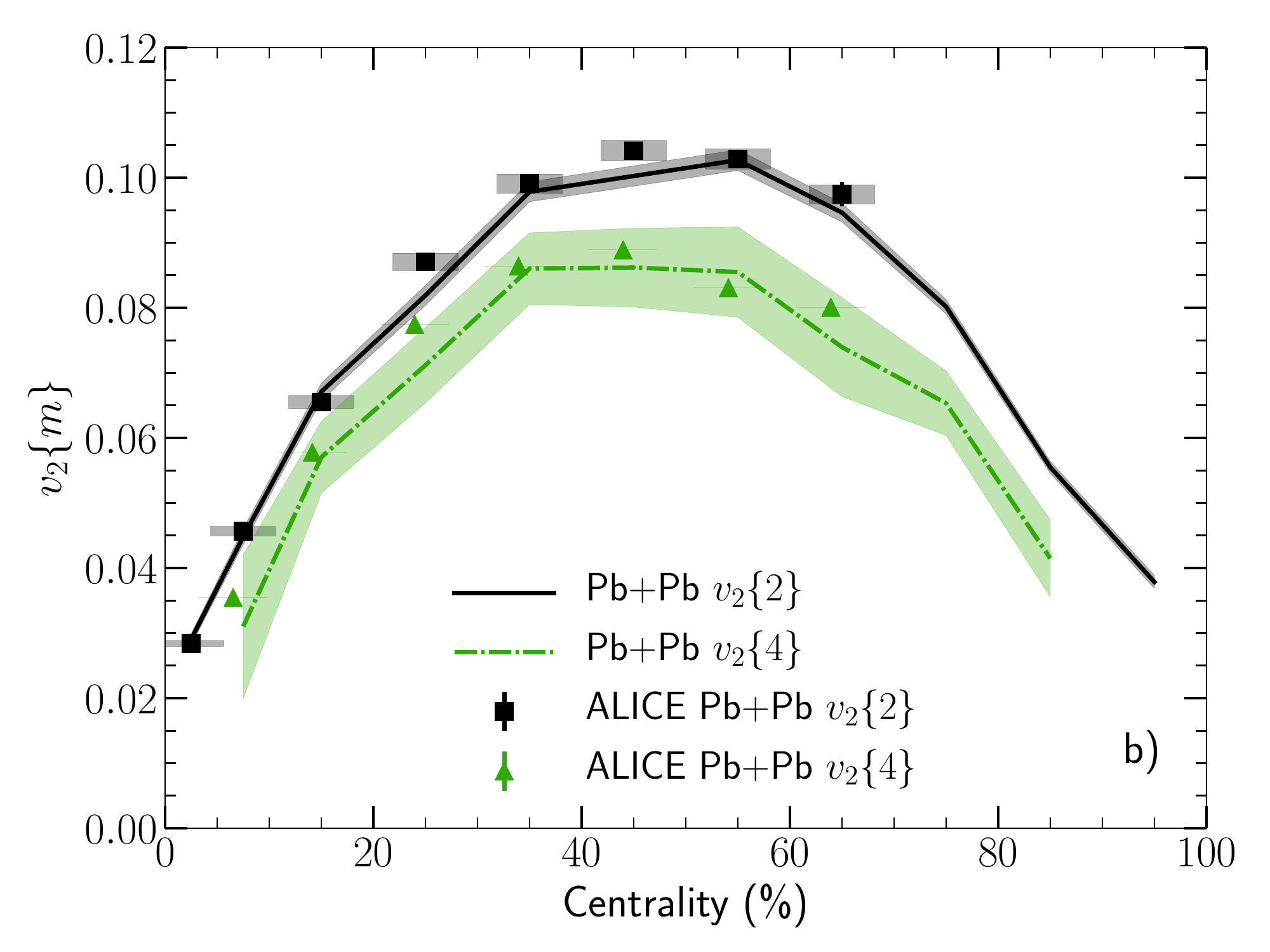}
  \caption{Elliptic anisotropies $v_2\{2\}$ and $v_2\{4\}$ for charged hadrons vs.~centrality in Au+Au collisions at RHIC (a) and Pb+Pb collisions at LHC (b). Experimental data from STAR \cite{Braidot:2009ji} and ALICE \cite{Adam:2016izf}. \label{fig:v2}}
\vspace{-0.3cm}
\end{figure}

In Fig.\,\ref{fig:mult} we present charged hadron and identified particle multiplicities  as functions of centrality at RHIC and LHC energies, including predictions for O+O collisions at LHC. In Fig.\,\ref{fig:pt} we show the $\langle p_T\rangle$ of identified particles in Au+Au collisions at RHIC and Pb+Pb collisions at the LHC. Agreement with the experimental data is good, with our proton $\langle p_T\rangle$ being slightly larger than the data at RHIC, but smaller for most centralities at LHC. Results for $v_2\{2\}$ and $v_2\{4\}$ in Au+Au collisions at RHIC and Pb+Pb collisions at LHC are shown in Fig.\,\ref{fig:v2}. Agreement with the data is very good, with expected deviations occurring for $v_2\{2\}$ in peripheral events, where non-flow effects (not included in the calculation) have a sizable contribution. 

\begin{figure}[tb]
  \begin{center}
  \includegraphics[width=0.47\textwidth]{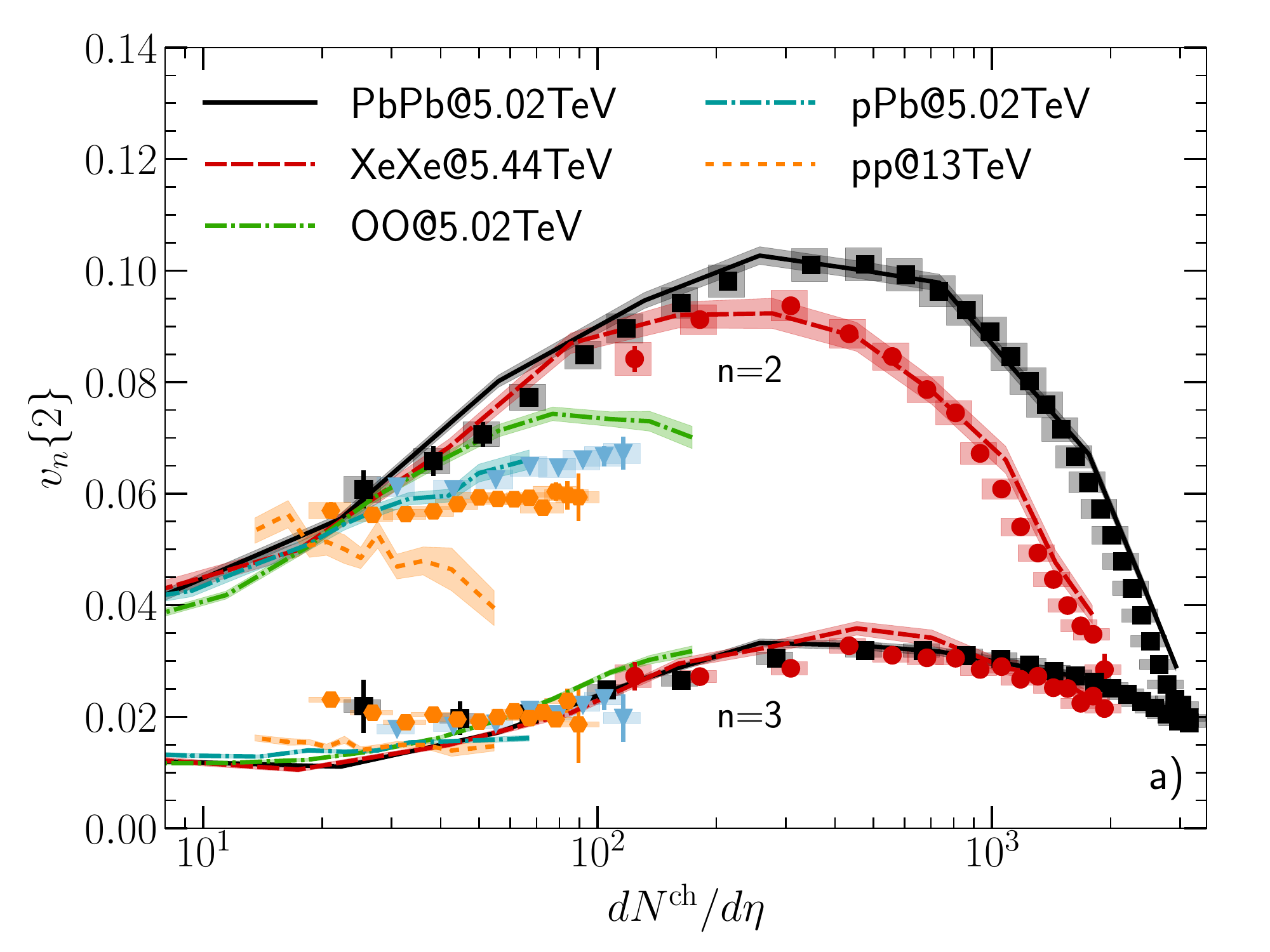}\includegraphics[width=0.47\textwidth]{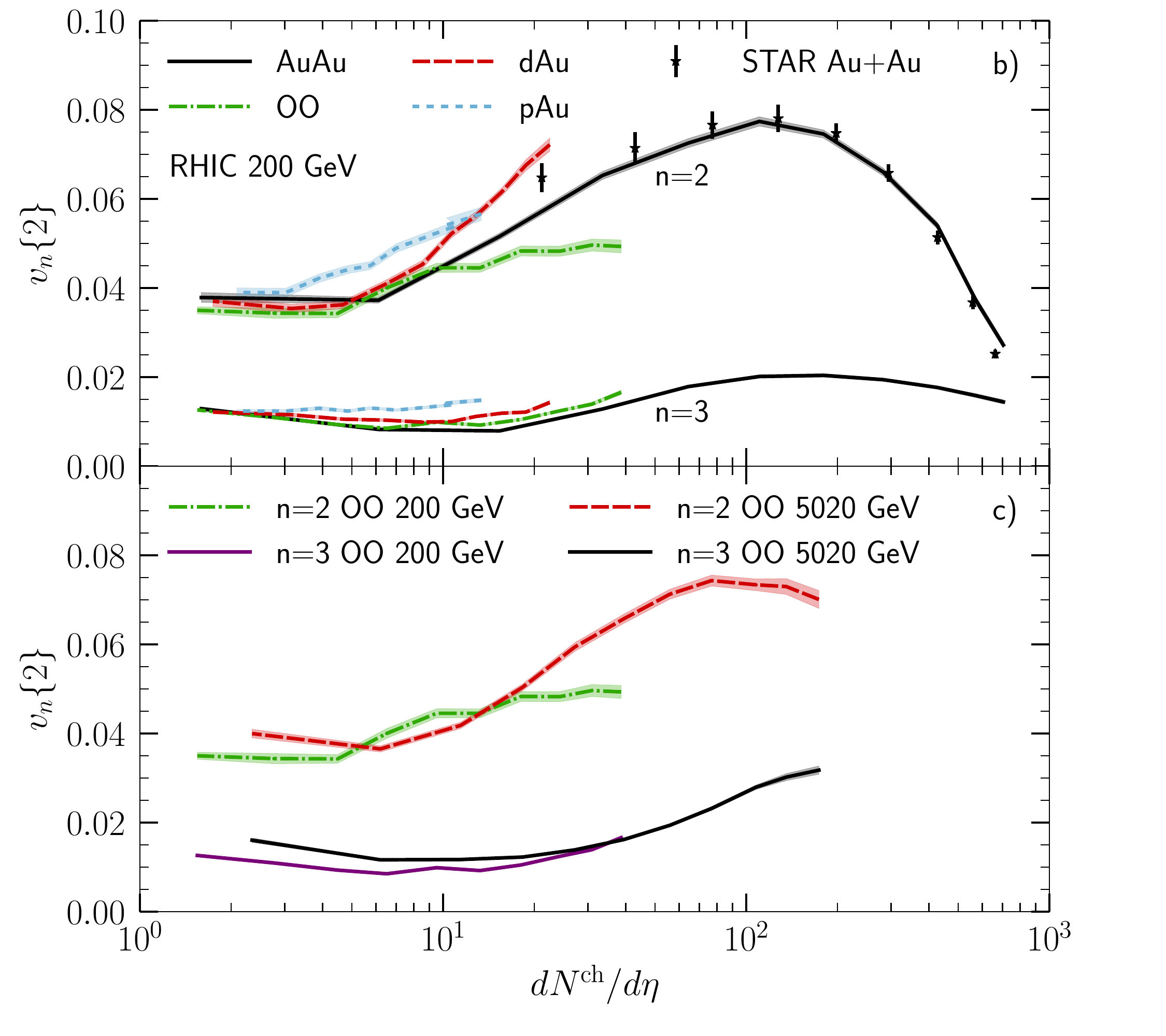}
  \caption{Elliptic anisotropies $v_2\{2\}$ and $v_3\{2\}$ for charged hadrons vs.~centrality at LHC (a), and RHIC (b), and comparison of predictions of $v_n\{2\}$ in O+O at RHIC and LHC (c). Experimental data from STAR \cite{Adam:2019woz} and ALICE \cite{Acharya:2019vdf}. \label{fig:v2OO}}
  \end{center}
\vspace{-0.3cm}
\end{figure}

In Fig.\,\ref{fig:v2OO} a) we present the comparison of calculated $v_2\{2\}$ and $v_3\{2\}$ vs. $dN_{\rm ch}/d\eta$ in different collision systems at LHC to experimental data. Agreement is very good down to small multiplicities and small system sizes, except for pp collisions, where we see a decreasing trend of $v_2\{2\}$ with multiplicity, compared to a slight increase in the experimental data. This could indicate that final state effects, that typically lead to an increasing $v_2$ with multiplicity, are underestimated in the calculation, relative to initial state anisotropies, which decrease with multiplicity \cite{Schenke:2019pmk}. Another source of discrepancy could be the method of binning and averaging in multiplicities. 
Fig.\,\ref{fig:v2OO} b) shows $v_n\{2\}$ for different collision systems at RHIC. We note that we predict $v_n\{2\}$(p/dAu 200 GeV)$>$ $v_n\{2\}$(AuAu 200 GeV), while $v_n\{2\}$(pPb 5020 GeV)$<$ $v_n\{2\}$(PbPb 5020 GeV) at the respective highest multiplicity reached in the smaller system. Predictions for $v_2\{2\}$ and $v_3\{2\}$ as functions of multiplicity in O+O collisions at 200\,GeV and 5.02\,TeV are presented in Fig.\,\ref{fig:v2OO} c). 

\section{Conclusions}
We have demonstrated that a comprehensive description of bulk observables and multiparticle correlations in a large variety of collision systems at top RHIC energies and above is possible within the hybrid model consisting of IP-Glasma, \textsc{Music}, and UrQMD. Having established this, we can use the model to make predictions for other systems and energies (done here for O+O), more complex observables, and extract information on the initial state and transport properties of the quark gluon plasma.

\section*{ Acknowledgments} 
B.P.S. and P.T. are supported under DOE Contract No. DE-SC0012704. C.S. is supported under DOE Contract No. DE-SC0013460. This research used resources of the National Energy Research Scientific Computing Center, which is supported by the Office of Science of the U.S. Department of Energy under Contract No. DE-AC02-05CH11231 and resources of the high performance computing services at Wayne State University. This work is supported in part by the U.S. Department of Energy, Office of Science, Office of Nuclear Physics, within the framework of the Beam Energy Scan Theory (BEST) Topical Collaboration.





\bibliographystyle{elsarticle-num}
\bibliography{spires}







\end{document}